\documentclass[reprint, superscriptaddress, amsmath, amssymb, aps, prb]{revtex4-2}
\usepackage[dvipdfmx]{graphicx}
\usepackage{dcolumn}
\usepackage{bm}
\usepackage[dvipdfmx]{hyperref}
\usepackage{comment}
\usepackage{color}
\usepackage{physics}
\begin{document}
\preprint{APS/123-QED}
\title{Majorana flat bands and anomalous proximity effects \\ in $p$-wave magnet--superconductor hybrid systems}
\author{Yutaro Nagae}
\thanks{These authors contributed equally to this work.}
\affiliation{Department of Applied Physics, Nagoya University, Nagoya 464-8603, Japan}
\author{Leo Katayama}
\affiliation{Department of Applied Physics, Nagoya University, Nagoya 464-8603, Japan}
\author{Satoshi Ikegaya}
\thanks{These authors contributed equally to this work.}
\affiliation{Department of Applied Physics, Nagoya University, Nagoya 464-8603, Japan}
\affiliation{Institute for Advanced Research, Nagoya University, Nagoya 464-8601, Japan}
\date{\today}

\begin{abstract}
Flat-band Majorana bound states of nodal $p$-wave superconductors give rise to striking electromagnetic anomalies, reflecting their high degree of degeneracy at the Fermi level.
However, experimental investigations of these states have been limited because of the scarcity of materials exhibiting intrinsic $p$-wave superconductivity.
In this Letter, we demonstrate that Majorana flat bands can emerge in a hybrid system consisting of a conventional superconductor and a $p$-wave magnet,
a recently proposed class of unconventional magnets that possess a unique composite symmetry, the $[C_{2\perp}||\boldsymbol{t}]$ symmetry.
The degeneracy of the flat-band Majorana bound states is protected by chiral symmetry from the BDI symmetry class,
which originates from the $[C_{2\perp}||\boldsymbol{t}]$ symmetry of the $p$-wave magnet.
In addition, we predict the robust appearance of a zero-bias conductance peak in a dirty normal-metal--superconductor junction containing a $p$-wave magnet,
which serves as an unambiguous signature of anomalous proximity effects associated with the Majorana flat bands.
\end{abstract}
\maketitle

\textit{Introduction.}
Majorana bound states (MBSs) in topological superconductors (SCs) have been a central focus in condensed matter physics over the past two decades%
~\cite{kane_10,zhang_11,nagaosa_12,beenakker_15,sato_17}.
Extensive studies have shown that MBSs manifest in various forms depending on the topological nature of the parent SC.
In two dimensions, for example, a chiral $p$-wave SC in class D of the Altland--Zirnbauer symmetry classification hosts a dispersing MBS (chiral MBS)~\cite{volovik_97,green_00,furusaki_01},
while a helical $p$-wave SC in class DIII harbors a pair of MBSs moving in opposite directions (helical MBSs)~\cite{schnyder_08,zhang_09}.
In addition, a nodal $p$-wave SC supports topologically protected flat-band MBSs (FMBSs)~\cite{nagai_86,sarma_01,tanuma_01,sato_11}, which are the focus of this Letter.
The FMBSs are distinguished by their exceptionally high degree of degeneracy at the Fermi level, which results in striking transport anomalies,
such as robust zero-bias conductance peaks in dirty normal-metal--superconductor (DN--SC) junctions~\cite{tanaka_04,tanaka_05(1),asano_07,ikegaya_15}
and fractional Josephson effects in SC--DN--SC junctions~\cite{asano_06(1),asano_06(2),ikegaya_16(2)}.
These phenomena are collectively referred to as anomalous proximity effects.

Experimental studies of FMBSs have been severely limited by the scarcity of materials exhibiting intrinsic $p$-wave superconductivity.
To overcome this challenge, several theoretical models have been proposed to realize effective nodal $p$-wave superconductivity,
including spin-singlet $s$- or $d$-wave superconductors in coexistence with a persistent spin-helix~\cite{alicea_10,you_13,lee_21,ikegaya_21,ikegaya_22,lee_25},
spin-singlet $s$-wave superconductors coupled to helical spins~\cite{nagaosa_13,schnyder_15,bena_15,chatterjee_24},
and Ising superconductors under an in-plane magnetic field~\cite{law_18,ojanen_20}.
Despite these theoretical proposals, experimental realizations of FMBSs in these systems have not yet been achieved.
Thus, investigating alternative pathways for realizing FMBSs continues to be a crucial and open area of research.

\begin{figure}[b]
\begin{center}
\includegraphics[width=0.5\textwidth]{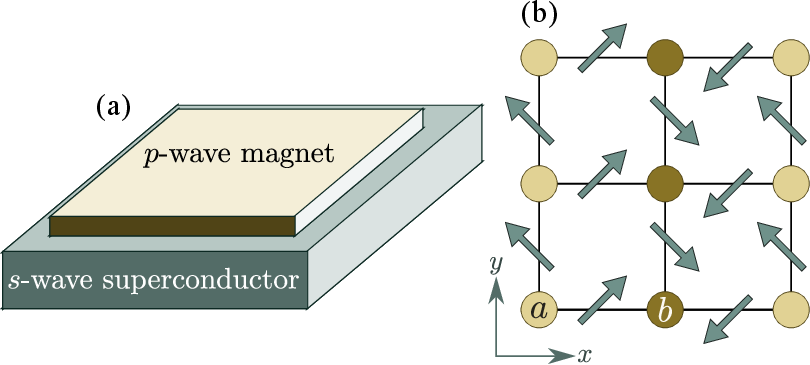}
\caption{(a) Schematic of the $p$M--SC heterostructure.
(b) Schematic of the lattice model with coplanar non-collinear spins that give rise to $p$-wave magnetism.
The spin-dependent hopping integral across the magnetic atoms, with the magnetic moments oriented along the $\pm (x+y)$-direction ($\pm (-x+y)$-direction),
is coupled with $\pm s_{x}$ ($\pm s_{y}$) in Eq.~(\ref{eq:nmham}).
}
\label{fig:figure1}
\end{center}
\end{figure}
\begin{figure}[t]
\begin{center}
\includegraphics[width=0.4\textwidth]{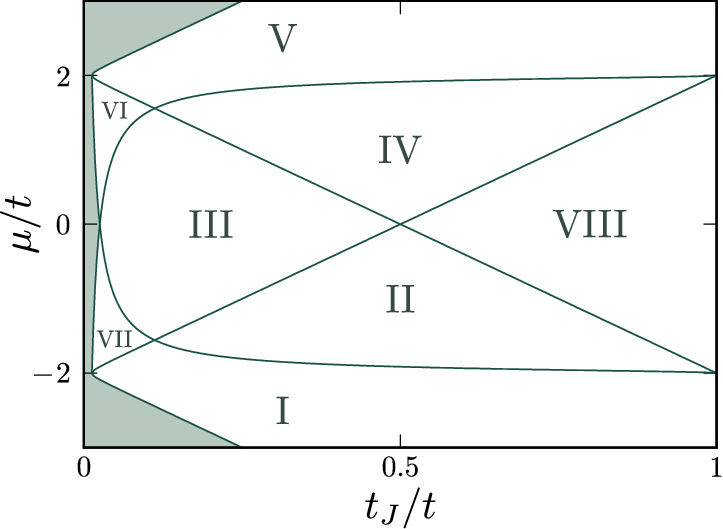}
\caption{Topological phase diagram of the $p$M--SC heterostructure for $\Delta=0.05t$.
There are eight distinct nodal topological phases, I--VIII, each hosting FMBSs.
The gray-shaded region corresponds to a fully gapped non-topological phase.
The phases are distinguished by the number of superconducting gap nodes on the Fermi surfaces.}
\label{fig:figure2}
\end{center}
\end{figure}
\begin{figure*}[t]
\begin{center}
\includegraphics[width=0.8\textwidth]{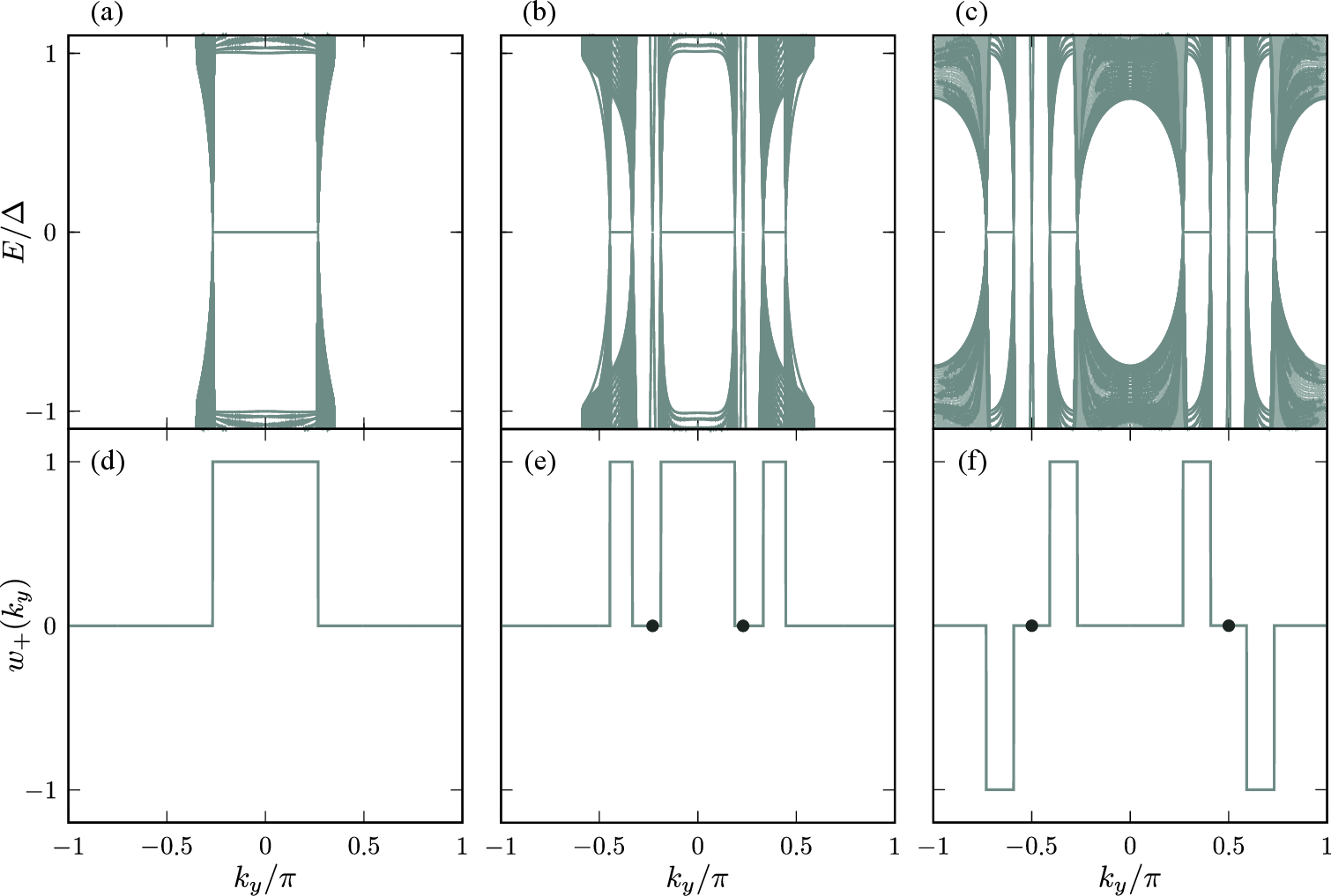}
\caption{Energy spectra and winding numbers of the nodal topological phases.
The upper panels show the energy spectra for phases (a) I, (b) II, and (c) III; the corresponding winding numbers $w_+(k_y)$ are plotted in lower panels (d)--(f).
The presence of FMBSs is confirmed in the momentum range where $w_+(k_y)\neq0$.
The black dots in (e) and (f) indicate the momenta where additional band crossings associated with accidental edge states occur.}
\label{fig:figure3}
\end{center}
\end{figure*}
\begin{figure*}[t]
\begin{center}
\includegraphics[width=0.9\textwidth]{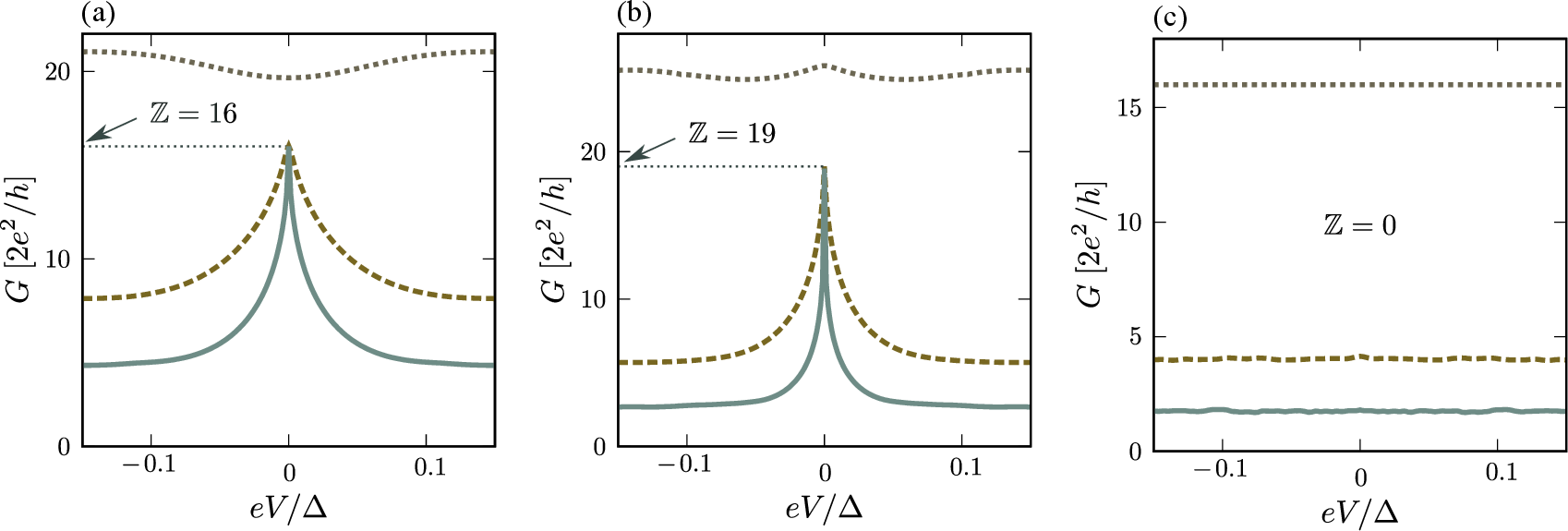}
\caption{Differential conductance as a function of the bias voltage for phases (a) I, (b) II, and (c) III.
The dotted, dashed, and solid lines represent the results for $v_{\mathrm{imp}}=0.25t$, $1.0t$, and $1.5t$, respectively.
In (a) and (b), a zero-bias conductance peak is observed for stronger disorder ($v_{\mathrm{imp}}=1.0t$ and $1.5t$), with the zero-bias conductance quantized as $G(0)=(2e^2/h) \mathbb{Z}$.
In (c), the zero-bias conductance peak is absent because $\mathbb{Z}=0$ for the chosen parameters.
}
\label{fig:figure4}
\end{center}
\end{figure*}

In this Letter, we demonstrate the realization of FMBSs in a hybrid system consisting of a spin-singlet $s$-wave SC and a $p$-wave magnet ($p$M)---
a recently proposed material exhibiting a unique coplanar non-collinear magnetic order~\cite{smejkal_23,smejkal_24,agterberg_25}---as shown in Fig.~\ref{fig:figure1}.
The interplay between superconductivity and unconventional magnetism, including altermagnetism%
~\cite{seo_19,hayami_19,hayami_20,smejkal_20,tsymbal_21,smejkal_22(0),smejkal_22(1),smejkal_22(2),mazin_22}
and $p$-wave magnetism, is currently a highly active area of research%
~\cite{linder_23,papaj_23, soori_24,wang_24,neupert_23,linder_23(2),zyuzin_24,belzig_24,linder_23(3),beenakker_23,sun_23,%
tanaka_24,scheurer_24,sum_24,linder_24(2),yan_23,liu_23,sudbo_23,cano_23,ikegaya_24,nag_24,saha_24, maeda_24,fukaya_24,linder_24,maeda_25,soori_25},
with recent theoretical investigations proposing the possibility of topological superconductivity in one-dimensional $p$M--SC hybrid systems~\cite{law_25,ezawa_24}.
Here, we show that, in two dimensions, the combination of $s$-wave superconductivity and $p$-wave magnetism gives rise to nodal topological superconductivity that supports FMBSs.
Importantly, $p$Ms inherently preserve a composite time-reversal-like symmetry originating from their characteristic spin arrangement~\cite{smejkal_23}.
In the superconducting states, this time-reversal-like symmetry induces chiral symmetry belonging to the BDI symmetry class,
which is essential for stabilizing FMBSs against surface roughness~\cite{ikegaya_17,ikegaya_18}.
We demonstrate the realization of the nodal topological phases in a broad parameter space (Fig.~\ref{fig:figure2})
and confirm the emergence of the FMBSs at the system edges (Fig.~\ref{fig:figure3}).
In addition, we investigate the transport properties of $p$M-based DN--SC junctions using recursive Green's function techniques~\cite{fisher_81,ando_91}.
We observe the robust appearance of the zero-bias conductance peak, associated with an Atiyah--Singer index~\cite{ikegaya_16(1)},
which serves as a hallmark of the anomalous proximity effect induced by the FMBSs (Fig.~\ref{fig:figure4}).
Our findings offer a promising new avenue for observing the striking phenomena induced by FMBSs.

\textit{Majorana flat bands.}
We consider a $p$M--SC heterostructure as shown in Fig.~\ref{fig:figure1}.
To describe the $p$M, we use a two-dimensional minimal model originally proposed in Ref.~[\onlinecite{smejkal_23}].
Although we focus on a two-dimensional $p$M in this study,
our main conclusions remain applicable to systems in which $p$M layers are stacked along the $z$-direction, provided that all layers exhibit the proximity-induced superconducting gap. 
The Bogoliubov--de Gennes (BdG) Hamiltonian is given by
\begin{align}
\begin{split}
&H=\frac{1}{2}\sum_{\boldsymbol{k}} [c^{\dagger}_{\boldsymbol{k}},c^{\mathrm{T}}_{-\boldsymbol{k}}]
H(\boldsymbol{k}) \left[ \begin{array} {cc} c_{\boldsymbol{k}}\\ c^{\ast}_{-\boldsymbol{k}} \end{array} \right],\\
&c^{\dagger}_{\boldsymbol{k}}=[ e^{-i \frac{k_x}{2}}c^{\dagger}_{\boldsymbol{k},a}, e^{i \frac{k_x}{2}}c^{\dagger}_{\boldsymbol{k},b}],\quad
c^{\dagger}_{\boldsymbol{k},\alpha}=[c^{\dagger}_{\boldsymbol{k},\alpha,\uparrow},c^{\dagger}_{\boldsymbol{k},\alpha,\downarrow}],\\
& H(\boldsymbol{k})=\left[ \begin{array}{cc} h(\boldsymbol{k}) & \Delta (i s_y) \\ -\Delta (i s_y) & -h^{\ast}(-\boldsymbol{k}) \end{array}\right],
\label{eq:bdgham}
\end{split}
\end{align}
with the single-particle Hamiltonian
\begin{align}
h(\boldsymbol{k}) =& -2t (\cos \frac{k_x}{2} \rho_x + \cos k_y  ) - \mu \nonumber \\
&-2t_J (\sin \frac{k_x}{2} s_x \rho_y + \cos k_y s_y \rho_z ),
\label{eq:nmham}
\end{align}
where $c_{\boldsymbol{k},\alpha,s}$ is the annihilation operator for an electron at sublattice $\alpha=a,b$ and with momentum $\boldsymbol{k}$ and spin $s=\uparrow, \downarrow$.
The Pauli matrices $\rho_{\nu}$ and $s_{\nu}$ for $\nu=x,y,z$ denote the sublattice and spin degrees of freedom, respectively.
The nearest-neighbor hopping integral and chemical potential are given by $t$ and $\mu$, respectively.
The spin-dependent hopping integral across the magnetic atoms, which gives rise to the $p$-wave magnetism, is represented by $t_J$ [Fig.~\ref{fig:figure1}(b)].
The proximity-induced spin-singlet $s$-wave pair potential is denoted by $\Delta$.

To define the topological invariant, we examine the symmetry of $H(\boldsymbol{k})$.
The $p$M generally preserves the $[C_{2\perp}||\boldsymbol{t}]$ symmetry~\cite{smejkal_23}, where
$C_{2\perp}$ denotes a $\pi$ spin rotation around an axis perpendicular to the spins and $\boldsymbol{t}$ represents a translation by half a unit cell.
For the present Hamiltonian, this symmetry is explicitly represented by
\begin{align}
\begin{split}
&[H(\boldsymbol{k}),U_p]=0, \quad U_p= U_{C_{2\perp}} U_{\boldsymbol{t}},\\
&U_{C_{2\perp}} = i s_z \tau_z, \quad U_{\boldsymbol{t}} = e^{-i\frac{k_x}{2}} \rho_x,
\end{split}
\end{align}
where $\tau_{\nu}$ ($\nu=x,y,z$) are the Pauli matrices in Nambu space.
Because of the magnetism, $H(\boldsymbol{k})$ does not preserve the usual time-reversal symmetry $T = i s_y \mathcal{K}$, where $\mathcal{K}$ denotes the complex conjugation operator.
However, the $p$-wave magnetic order induces time-reversal-like symmetries:
\begin{align}
\begin{split}
&T_{\pm} H(\boldsymbol{k}) T^{-1}_{\pm} = H(-\boldsymbol{k}),\\
&T_+ = U_{C_{2\perp}} T, \quad T_- = U_{\boldsymbol{t}}T.
\end{split}
\end{align}
Notably, while the original paper proposing the $p$M focuses on the emergence of the $T_-$ symmetry with $T_-^2=-1$~\cite{smejkal_23},
the $p$-wave magnetism simultaneously guarantees the presence of the $T_+$ symmetry, where $T_+^2=1$.
The BdG Hamiltonian also inherently preserves particle-hole symmetry:
\begin{align}
C H(\boldsymbol{k}) C^{-1} = - H(-\boldsymbol{k}), \quad C=\tau_x \mathcal{K}.
\end{align}
Thus, in terms of $T_+$ ($T_-$), the BdG Hamiltonian $H(\boldsymbol{k})$ belongs to class BDI (DIII) of the Altland--Zirnbauer symmetry classification~\cite{schnyder_08}.
By combining the time-reversal-like symmetry and particle-hole symmetry, we obtain chiral symmetry:
\begin{align}
\begin{split}
&\Gamma_{\pm} H(\boldsymbol{k}) \Gamma_{\pm}^{-1} = - H(\boldsymbol{k}),\\
&\Gamma_+ = T_+ C, \quad \Gamma_- = i T_- C.
\end{split}
\end{align}
Using the chiral symmetry, we can define the one-dimensional winding number for each $k_y$ as,
\begin{align}
w_{\pm}(k_y) = \frac{i}{4 \pi} \int dk_x \mathrm{Tr}[\Gamma_{\pm} H^{-1}(\boldsymbol{k}) \partial_{k_x} H(\boldsymbol{k}) ],
\label{eq:wind_num}
\end{align}
which topologically characterizes the FMBSs that appear at the edges perpendicular to the $x$-axis~\cite{sato_11}.
Notably, the winding number $w_{\pm}(k_y)$, which depends on the momentum $k_y$,
predicts the existence of the MBSs only at \emph{clean} edges, where $k_y$ remains a good quantum number.
For \emph{dirty} edges, the number of MBSs is instead determined by an Atiyah--Singer index defined using the winding number of the class BDI~\cite{ikegaya_17,ikegaya_18}:
\begin{align}
\mathbb{Z}={\sum_{k_y}}^{\prime} w_{+}(k_y),
\label{eq:as_index}
\end{align}
where $\sum^{\prime}_{k_y}$ denotes a summation over $k_y$ excluding nodal points.

Before explicitly calculating the winding number, we derive a band-basis Hamiltonian, which provides an intuitive understanding of the emergence of the FMBSs.
We first define a unitary matrix that diagonalizes the $[C_{2\perp}||\boldsymbol{t}]$ symmetry operator:
\begin{align}
A U_p A^{\dagger} = i e^{-i\frac{k_x}{2}} \left[ \begin{array}{cc} 1 & 0 \\ 0 & -1 \end{array} \right],
\end{align}
where the eigenvalues of $U_p$ are given by $\lambda_{\pm}(k_x) = \pm i e^{-i\frac{k_x}{2}}$.
Because $H(\boldsymbol{k})$ commutes with $U_p$, we can decompose the Hamiltonian into two subsectors, each corresponding to a different eigenvalue of $U_p$:
\begin{align}
A H(\boldsymbol{k}) A^{\dagger} =  \left[ \begin{array}{cc} H_+(\boldsymbol{k}) & 0 \\ 0 & H_-(\boldsymbol{k}) \end{array} \right],
\label{eq:band_ham}
\end{align}
where $H_{\pm}(\boldsymbol{k})$ is associated with the eigenvalue $\lambda_{\pm}(k_x)$.
The explicit form of the unitary matrix $A$ is provided in the Supplemental Material (SM)~\cite{sm}.
Because $\lambda_{\pm}(k_x+2\pi)=\lambda_{\mp}(k_x)$, we obtain the relation $H_{\pm}(k_x+2\pi,k_y)=H_{\mp}(\boldsymbol{k})$~\cite{kane_15}.
On the basis of this relation, we define $\mathcal{H}(p_x,k_y)=H_{\pm}(p_x \pm \pi, k_y)$, which has the explicit form
\begin{align}
\mathcal{H}(p_x,k_y) =
\left[ \begin{array}{cc} \mathcal{H}^{\mathrm{band}}_+(p_x,k_y) & -i\mathcal{V}(p_x,k_y) \sigma_x \\ i\mathcal{V}(p_x,k_y) \sigma_x & \mathcal{H}^{\mathrm{band}}_-(p_x,k_y) \end{array} \right],
\end{align}
where
\begin{align}
\begin{split}
&\mathcal{H}_{\pm}^{\mathrm{band}} = \xi_{\pm} \sigma_z \pm \Delta_{\mathrm{eff}} \sigma_x,\\
&\xi_{\pm}=-2t \cos k_y \pm \mathcal{M} - \mu, \quad 
\Delta_{\mathrm{eff}}=\frac{2t \Delta}{\mathcal{M}} \sin \frac{p_x}{2},\\
&\mathcal{V} = \frac{2t_J \Delta}{\mathcal{M}} \left( \cos \frac{p_x}{2} + \cos k_y \right),\\
&\mathcal{M} = 2 \sqrt{t^2 \sin^2 \frac{p_x}{2} + t_J^2 \left( \cos \frac{p_x}{2} + \cos k_y \right)^2},
\end{split}
\end{align}
and $\sigma_{\nu}$ ($\nu=x,y,z$) are the Pauli matrices.
The block component $\mathcal{H}_{\pm}$ corresponds to the BdG Hamiltonian with respect to the band described by $\xi_{\pm}$,
while the off-diagonal component $\mathcal{V}$ describes the additional band hybridization induced by the pair potential.
Importantly, the effective pair potential acting on each band exhibits nodal $p$-wave pairing symmetry:
\begin{align}
\Delta_{\mathrm{eff}}(p_x,k_y)=-\Delta_{\mathrm{eff}}(-p_x,k_y)=\Delta_{\mathrm{eff}}(p_x,-k_y).
\label{eq:eff_pair}
\end{align}
As a result, we can expect the emergence of FMBSs in the $p$M--SC heterostructure.

We now describe the topological phase diagram of the present system.
As detailed in the SM~\cite{sm}, the winding number in Eq.~(\ref{eq:wind_num}) can be further simplified as follows:
\begin{align}
\begin{split}
&w_+(k_y) = \frac{1}{2}\left\{\mathrm{sgn}[R(2\pi,k_y)] - \mathrm{sgn}[R(0,k_y)]  \right\},\\
&R(p_x,k_y) = \xi_+(p_x,k_y) \xi_-(p_x,k_y) + \Delta^2,
\label{eq:wind_simplified}
\end{split}
\end{align}
and $w_{-}(k_y)=0$ identically.
Thus, the FMBSs in the present symmetry are characterized by the winding number in the class BDI.
The superconducting gap nodes on the Fermi surfaces appear at the momenta $(k_x,k_y) = (\pi,\pm k_n)$, where $k_n$ satisfies $R(0,\pm k_n)=0$ or $R(2\pi,\pm k_n)=0$.
On the basis of the number of superconducting gap nodes, we identify eight distinct nodal topological phases, as summarized in Fig.~\ref{fig:figure2}.
Phases I and V exhibit two nodes, phases VI, VII, and VIII have four nodes, phases II and IV display six nodes, and phase III has eight nodes.
The gray-shaded region corresponds to a fully gapped non-topological phase.
The explicit functions for the topological phase boundaries are provided in the SM~\cite{sm}; we use $\Delta=0.05t$ in Fig.~\ref{fig:figure2}.
We also note that phases VI and VII vanish in the limit of $\Delta/t \rightarrow 0$.
At $\mu/t=0$ ($\mu/t=\pm2$), for instance, the nodal topological phase is realized for $t_J > \Delta \sqrt{t^2/(4t^2+\Delta^2)}$ ($t_J > \Delta/4$).
Given that the spin splitting of the bands due to the $p$-wave magnetism is expected to reach up to 200 $\mathrm{meV}$~\cite{smejkal_23}, we anticipate $t_J \gg \Delta$.
Consequently, the realization of the nodal topological phases can be expected over a broad range of realistic $t_J$.
In the main text, we focus on the FMBSs in phases I, II, and III; those in the other phases are discussed in SM~\cite{sm}.
In Figs.~\ref{fig:figure3}(a)--\ref{fig:figure3}(c), we show the energy spectra of the present system with an open boundary condition in the $x$-direction.
For the numerical calculations, we rewrite the Hamiltonian by applying a Fourier transformation with respect to the $x$-direction:
$c_{\boldsymbol{k},\alpha,s} \rightarrow c_{i_x, k_y,\alpha,s}$, where $i_x$ denotes the position along the $x$-axis.
We assume that the open edges are located at $i_x=1$ and $i_x=400$.
To describe phases I, II, and III, we choose the parameters as $(\mu/t,t_J/t)=(-3,0.5)$, $(-1.5,0.5)$, and $(0, 0.4)$, respectively, where the pair potential is fixed at $\Delta=0.05t$.
In Figs.~\ref{fig:figure3}(d)--\ref{fig:figure3}(f), we plot the winding number $w_+(k_y)$ for these phases.
We clearly observe that the FMBSs appear in the momentum ranges where $w_+(k_y) \neq 0$.
In Figs.~\ref{fig:figure3}(b) and \ref{fig:figure3}(c), we also find additional band crossings between the flat-bands,
which occur specifically at $k_y=\pm \arccos(-\mu/2t)$, as indicated by the dots in Figs.~\ref{fig:figure3}(e) and \ref{fig:figure3}(f).
As detailed in the SM~\cite{sm}, these band crossings are due to additional edge states protected by an accidental chiral symmetry preserved \emph{only} at $k_y=\pm \arccos(-\mu/2t)$.
Most importantly, we observe a non-zero Atiyah--Singer index, i.e., $\mathbb{Z}=\sum^{\prime}_{k_y} w_{+}(k_y)\neq0$, as shown in Figs.~\ref{fig:figure3}(d) and \ref{fig:figure3}(e).
In Fig.~\ref{fig:figure3}(f), the momentum ranges for $w_+(k_y)=1$ and $w_+(k_y)=-1$ are of equal width, leading to $\mathbb{Z}=0$.
However, we confirm that this offsetting occurs only at $\mu=0$~\cite{sm}.
Therefore, the FMBSs in the present system, characterized by a non-zero $\mathbb{Z}$ (except at $\mu=0$), are robust against surface roughness.

\textit{Anomalous proximity effect.}
We here discuss the anomalous proximity effect of the $p$M--SC hybrids.
To study the transport properties, we consider a junction consisting of three segments:
a ballistic normal-metal segment for $-\infty \leq i_x < 1$, a DN region for $1 \leq i_x \leq L$, and a SC segment for $L<i_x \leq \infty$.
In the $y$-direction, we apply an open boundary condition, where the number of lattice sites is denoted by $W$.
For numerical calculations, we rewrite the BdG Hamiltonian in Eq.~(\ref{eq:bdgham}) in a real space basis by applying a Fourier transformation:
$c_{\boldsymbol{k},\alpha,s} \rightarrow c_{\boldsymbol{r},\alpha,s}$,
where $\boldsymbol{r} = i_x \boldsymbol{x} + i_y \boldsymbol{y}$, with $\boldsymbol{x}$ ($\boldsymbol{y}$) being the unit vector in the $x$-direction ($y$-direction).
We assume that the proximity-induced pair potential exists only in the SC segment, i.e., $i_x>L$.
For the DN segment, we introduce a non-magnetic random on-site potential:
\begin{align}
H_d = \sum_{i_x=1}^{L}\sum_{i_y, \alpha, s}v_{\boldsymbol{r}, \alpha}  c^{\dagger}_{\boldsymbol{r},\alpha,s}c_{\boldsymbol{r},\alpha,s},
\end{align}
where $v_{\boldsymbol{r}, \alpha}$ is given randomly in the range $-v_{\mathrm{imp}}/2 \leq v_{\boldsymbol{r}, \alpha} \leq v_{\mathrm{imp}}/2$.
In addition, to prevent numerical instability, we introduce an insignificant long-range hopping term in the $x$-direction:
$H^{\prime}=-t^{\prime}\sum_{\boldsymbol{r},\alpha,s}( c^{\dagger}_{\boldsymbol{r}+\boldsymbol{x},\alpha,s}c_{\boldsymbol{r},\alpha,s} + \mathrm{h.c.})$, where $t^{\prime} \ll t$.
We note that these additional terms, i.e., $H_d$ and $H^{\prime}$, do not break the chiral symmetry $\Gamma_+$, which protects the degeneracy of the FMBSs.
We calculate the differential conductance at zero temperature on the basis of the Blonder--Tinkham--Klapwijk formalism~\cite{klapwijk_82}:
\begin{align}
G(eV) = \frac{e^{2}}{h} \sum_{\zeta,\eta}
\left[ \delta_{\zeta,\eta} - \left| r^{ee}_{\zeta,\eta} \right|^{2}
+ \left| r^{he}_{\zeta,\eta} \right|^{2} \right]_{E=eV},
\end{align}
where $r^{ee}_{\zeta,\eta}$ and $r^{he}_{\zeta,\eta}$ denote a normal and an Andreev reflection coefficient at energy $E$, respectively.
The indices $\zeta$ and $\eta$ label the outgoing and incoming channels, respectively, in the ballistic normal-metal segment.
These reflection coefficients are calculated using recursive Green's function techniques~\cite{fisher_81,ando_91}.
In the following calculations, we fix the parameters as $L=60$, $W=60$, $\Delta=0.05t$, and $t^{\prime}=0.01t$.

In Fig.~\ref{fig:figure4}, we show the differential conductance as a function of the bias voltage for phases (a) I, (b) II, and (c) III.
To describe phases I, II, and III, we choose the parameters as $(\mu/t,t_J/t)=(-3,0.5)$, $(-1.5,0.5)$, and $(0, 0.4)$, respectively.
For these parameters, the Atiyah--Singer index is 16 in phase I, 19 in phase II, and zero in phase III.
To calculate $\mathbb{Z}$, we replace $k_y$ in Eq.~(\ref{eq:as_index}) with $k_n = n \pi/(W+1)$ for $n=1$ to $W$, in accordance with the open boundary condition in the $y$-direction.
The disorder strengths are chosen as $v_{\mathrm{imp}}=0.25t$, $1.0t$, and $1.5t$, with an ensemble average taken over 100 samples.
For weak disorder ($v_{\mathrm{imp}}=0.25t$), as indicated by the dotted lines, the conductance spectra show no noticeable features.
However, for stronger disorder ($v_{\mathrm{imp}}=1.0t$ and $1.5t$),
the conductance spectra for phases I and II exhibit a prominent zero-bias peak, as indicated by the dashed and solid lines in Figs.~\ref{fig:figure4}(a) and \ref{fig:figure4}(b).
Remarkably, the minimum value of the zero-bias conductance is quantized at $(2e^2/h) \mathbb{Z}$, independent of the disorder strength.
This conductance quantization suggests the presence of $\mathbb{Z}$-fold degenerate resonant states
resulting from the penetration of the robust FMBSs from the SC segment into the DN segment~\cite{tanaka_04}.
By contrast, in Fig.~\ref{fig:figure4}(c), the conductance spectra do not display a zero-bias peak because $\mathbb{Z}=0$ at $\mu=0$.
As a result, we demonstrate the robust emergence of the zero-bias conductance peak in the $p$M--SC hybrid system, which is a key manifestation of the anomalous proximity effect caused by FMBSs.

\textit{Discussion.}
The study of the interplay between $p$-wave magnetism and superconductivity is still in its early stages.
Thus, in this Letter, we qualitatively explore the essential features of the $p$M--SC hybrids using a two-dimensional minimal model.
However, further investigation using more realistic models, such as those that explicitly describe a $p$M--SC heterostructure through a three-dimensional microscopic framework, is desired for future studies.
Although this paper focuses on the emergence of the FMBSs, MBSs have also been shown to be inherently linked to exotic odd-frequency Cooper pairs~\cite{golubov_07,asano_13,tamura_19}.
Therefore, investigating the pairing functions in the present system would be an intriguing direction for future research.

In summary, we have theoretically demonstrated the appearance of FMBSs in the $p$M--SC hybrid system.
The degeneracy of the FMBSs is protected by the chiral symmetry from the BDI symmetry class, which originates from the $[C_{2\perp}||\boldsymbol{t}]$ symmetry of the $p$-wave magnet.
In addition, we have shown the emergence of anomalous proximity effects, which serve as a distinctive signature of robust FMBSs.
Our findings provide a promising platform for the observation of dramatic electromagnetic anomalies driven by FMBSs, characterized by an exceptionally high degree of degeneracy.

\begin{acknowledgments}
S.I. is supported by a Grant-in-Aid for JSPS Fellows (JSPS KAKENHI Grant No. JP22KJ1507) and a Grant-in-Aid for Early-Career Scientists (JSPS KAKENHI Grant No. JP24K17010).
Y.N. is supported by JST SPRING (Grant No. JPMJSP2125) and thanks the THERS Make New Standards Program for the Next Generation Researchers.
\end{acknowledgments}

\clearpage
\onecolumngrid
\begin{center}
  \textbf{\large Supplemental Material for \\ ``Majorana flat bands and anomalous proximity effects \\ in $p$-wave magnet--superconductor hybrid systems''}\\ \vspace{0.3cm}
Yutaro Nagae$^{1}$, Leo Katayama$^{1}$, and Satoshi Ikegaya$^{1,2}$\\ \vspace{0.1cm}
{\itshape $^{1}$Department of Applied Physics, Nagoya University, Nagoya 464-8603, Japan\\
$^{2}$Institute for Advanced Research, Nagoya University, Nagoya 464-8601, Japan}
\date{\today}
\end{center}

\section{Band-basis Hamiltonian}
In this section, we derive a band-basis Hamiltonian for the $p$-wave magnet--superconductor ($p$M--SC) hybrid system, given by Eq.~(10) in the main text.
We start with the Bogoliubov--de Gennes (BdG) Hamiltonian in Eq.~(1) of the main text:
\begin{align}
\begin{split}
& H(\boldsymbol{k})=\left[ \begin{array}{cc} h(\boldsymbol{k}) & h_\Delta \\ - h_{\Delta} & -h^{\ast}(-\boldsymbol{k}) \end{array}\right],\\
&h(\boldsymbol{k}) = \left[\begin{array}{cc}
(-2t \cos k_y -\mu)I_2 -2t_J \cos k_y s_y & -2t \cos \frac{k_x}{2} I_2 + 2 i t_J \sin \frac{k_x}{2} s_x\\
-2t \cos \frac{k_x}{2} I_2 - 2 i t_J \sin \frac{k_x}{2} s_x & (-2t \cos k_y -\mu)I_2 + 2t_J \cos k_y s_y \end{array} \right],\\
&h_{\Delta} = \left[\begin{array}{cc} \Delta (i s_y) & O_2 \\ O_2 & \Delta (i s_y) \end{array} \right],
\end{split}
\end{align}
where $I_i$ denotes the $(i \times i)$ identity matrix, $O_i$ represents the $(i \times i)$ null matrix,
and the Pauli matrices in spin space are denoted by $s_{\nu}$ for $\nu=x,y,z$.
The nearest-neighbor hopping integral and chemical potential are given by $t$ and $\mu$, respectively.
The spin-dependent hopping integral due to the $p$-wave magnetism is represented by $t_J$.
The proximity-induced spin-singlet $s$-wave pair potential is denoted by $\Delta$.
The BdG Hamiltonian preserves the $[C_{2\perp}||\boldsymbol{t}]$ symmetry, originating from the characteristic spin arrangement of the $p$M~\cite{smejkal_23}:
\begin{align}
\begin{split}
&[H(\boldsymbol{k}),U_p]=0, \quad U_p= U_{C_{2\perp}} U_{\boldsymbol{t}},\\
&U_{C_{2\perp}} = i \left[ \begin{array}{cccc} s_z & O_2 & O_2 & O_2 \\ O_2 & s_z & O_2 & O_2 \\ O_2 & O_2 & -s_z & O_2 \\ O_2 & O_2 & O_2 & -s_z \end{array} \right], \quad
U_{\boldsymbol{t}} = e^{-i\frac{k_x}{2}}  \left[ \begin{array}{cccc} O_2 & I_2 & O_2 & O_2 \\ I_2 & O_2 & O_2 & O_2 \\ O_2 & O_2 & O_2 & I_2 \\ O_2 & O_2 & I_2 & O_2 \end{array} \right].
\end{split}
\end{align}
The $[C_{2\perp}||\boldsymbol{t}]$ symmetry operator is diagonalized as,
\begin{align}
A_1 U_p A_1^{\dagger} = i e^{-i\frac{k_x}{2}} \left[ \begin{array}{cc} I_4 & O_4 \\ O_4 & -I_4 \end{array} \right],
\end{align}
with
\begin{align}
A_1 = \frac{1}{\sqrt{2}} \left[ \begin{array}{cccc} I_2 & O_2 & O_2 & O_2 \\ O_2 & O_2 & O_2 & I_2 \\ O_2 & I_2 & O_2 & O_2 \\ O_2 & O_2 & I_2 & O_2 \end{array} \right]
\left[ \begin{array}{cccc} s_z & I_2 & O_2 & O_2 \\ s_z & -I_2 & O_2 & O_2 \\ O_2 & O_2 & -s_z & -I_2 \\  O_2 & O_2 & -s_z & I_2 \end{array} \right].
\label{eq:unitary1_sp}
\end{align}
Since $H(\boldsymbol{k})$ commutes with $U_p$, we can block-diagonalize the BdG Hamiltonian as,
\begin{align}
\begin{split}
&A_1 H(\boldsymbol{k}) A_1^{\dagger} = \left[ \begin{array}{cc} H^{\prime}_+(\boldsymbol{k}) & O_4 \\ O_4 & H^{\prime}_-(\boldsymbol{k}) \end{array} \right],\\
&H^{\prime}_{\pm}(\boldsymbol{k})=\left[ \begin{array}{cc} h_{\pm}(\boldsymbol{k}) & \Delta (i \sigma_y) \\ -\Delta (i \sigma_y) & -h_{\mp}^{\ast}(-\boldsymbol{k}) \end{array}\right],\\
&h_{\pm}(\boldsymbol{k})=-2t (\pm \cos \frac{k_x}{2} \sigma_z + \cos k_y  ) -\mu +2t_J (\pm \sin \frac{k_x}{2} + \cos k_y) \sigma_y,
\end{split}
\end{align}
where $\sigma_{\nu}$ ($\nu=x,y,z$) are the Pauli matrices.
Since $\lambda_{\pm}(k_x+2\pi)=\lambda_{\mp}(k_x)$, we obtain the relation: $H^{\prime}_{\pm}(k_x+2\pi,k_y)=H^{\prime}_{\mp}(\boldsymbol{k})$.
Using a unitary operator,
\begin{align}
\begin{split}
&A = A_2 A_1,\\
&A_2 =  \left[ \begin{array}{cc} \Phi(k_x,k_y) & O_4 \\ O_4 & \Phi(k_x+2\pi,k_y) \end{array}\right],\\
&\Phi(\boldsymbol{k}) = \left[ \begin{array}{cccc} 1 & 0& 0& 0 \\ 0 & 0& 1& 0\\ 0& 1& 0& 0\\ 0& 0& 0& 1 \end{array}\right]
\left[\begin{array}{cccc} \alpha & -i\beta & 0& 0\\ i\beta & -\alpha & 0& 0 \\ 0& 0& -i\beta & \alpha \\ 0& 0& -\alpha & i\beta \end{array} \right],\\
&\alpha = \frac{M-2t \cos (k_x/2)}{\sqrt{2 M \{ M-2t \cos (k_x/2)\}}}, \quad
\beta = \frac{2t_J \{ \sin (k_x/2) + \cos k_y \} }{\sqrt{2 M \{M-2t \cos (k_x/2)\}}}, \\
&M = 2 \sqrt{t^2 \cos^2 \frac{k_x}{2} + t_J^2 \left( \sin \frac{k_x}{2} + \cos k_y \right)^2},
\end{split}
\end{align}
we eventually obtain the band-basis Hamiltonian as,
\begin{align}
\begin{split}
&A H(\boldsymbol{k}) A^{\dagger} = \left[ \begin{array}{cc} H_+(\boldsymbol{k}) & O_4 \\ O_4 & H_-(\boldsymbol{k}) \end{array} \right],\\
&H_+(k_x,k_y)=H_-(k_x-2\pi,k_y)
= \left[ \begin{array}{cc} H^{\mathrm{band}}_+(\boldsymbol{k})  & -iV(\boldsymbol{k})  \sigma_x \\ iV(\boldsymbol{k})  \sigma_x & H^{\mathrm{band}}_-(\boldsymbol{k})  \end{array} \right],\\
&H_{\pm}^{\mathrm{band}} = \xi^{\prime}_{\pm} \sigma_z \pm \Delta^{\prime}_{\mathrm{eff}} \sigma_x,\\
&\xi^{\prime}_{\pm}=-2t \cos k_y \pm M - \mu, \quad 
\Delta^{\prime}_{\mathrm{eff}}=-\frac{2t \Delta}{M} \cos \frac{k_x}{2},\\
&V = \frac{2t_J \Delta}{M} \left( \sin \frac{k_x}{2} + \cos k_y \right),
\end{split}
\end{align}
which is equivalent to Eq.~(10) in the main text.

\section{Winding number and Topological Phase Diagram}
\subsection{Winding number for Majorana flat bands}
In this section, we derive a useful expression for the winding number, given in Eq.~(14) of the main text.
We first focus on the winding number $w_+(k_y)$, which belongs to class BDI:
\begin{align}
w_{+}(k_y) = \frac{i}{4 \pi} \int^{\pi}_0 dk_x \mathrm{Tr}[\Gamma_{+} H^{-1}(\boldsymbol{k}) \partial_{k_x} H(\boldsymbol{k}) ],
\end{align}
where
\begin{align}
\begin{split}
&\Gamma_{+} H(\boldsymbol{k}) \Gamma_{+}^{-1} = - H(\boldsymbol{k}),\quad
\Gamma_+ = U_{C_{2\perp}} T C, \\
 &T = i s_y \mathcal{K}, \quad C=\tau_x \mathcal{K}.
\end{split}
\end{align}
Here, $\mathcal{K}$ denotes the complex conjugation operator.
Using the unitary operator in Eq.~(\ref{eq:unitary1_sp}), we deform the winding number as,
\begin{align}
\begin{split}
w_{+}(k_y) &= \frac{i}{4 \pi} \int^{\pi}_0 dk_x \mathrm{Tr}[A_1\Gamma_{+}A_1^{\dagger} A_1 H^{-1}(\boldsymbol{k}) A_1^{\dagger} \partial_{k_x} A_1 H(\boldsymbol{k}) A_1^{\dagger}]\\
&=\frac{i}{4 \pi}\sum_{s=\pm} \int^{\pi}_0 dk_x \mathrm{Tr}[\gamma \{H_s^{\prime}(\boldsymbol{k})\}^{-1} \partial_{k_x} H^{\prime}_s (\boldsymbol{k}) ],
\end{split}
\end{align}
where
\begin{align}
\gamma = \left[ \begin{array}{cc} O_2 & i \sigma_x \\ -i \sigma_x & O_2 \end{array} \right].
\end{align}
Defining $\mathcal{H}^{\prime}(p_x,k_y)=H^{\prime}_{\pm}(p_x \pm \pi, k_y)$, we can rewrite the winding number:
\begin{align}
w_{+}(k_y)=\frac{i}{4 \pi}\int^{2\pi}_0 dp_x \mathrm{Tr}[\gamma \mathcal{H}^{\prime}(p_x,k_y) \partial_{p_x} \mathcal{H}(p_x,k_y) ].
\end{align}
To proceed with the calculation, we derive the chiral basis Hamiltonian:
\begin{align}
\begin{split}
&U_{\gamma} \mathcal{H}^{\prime}(p_x,k_y) U^{\dagger}_{\gamma} = \left[ \begin{array}{cc} O_2 & q(p_x,k_y) \\ q^{\dagger}(p_x,k_y) & O_2 \end{array} \right],\\
&U_{\gamma} = \frac{1}{\sqrt{2}}\left[ \begin{array}{cc} I_2 & -i \sigma_x \\ i \sigma_x & -I_2 \end{array} \right],\\
&q(p_x,k_y)=i (2t \cos k_y + \mu) \sigma_x + \left\{2t \sin \frac{p_x}{2} - i \Delta \right\} \sigma_y - 2 t_J \left\{ \cos k_y + \frac{p_x}{2} \right\} \sigma_z,\\
\end{split}
\end{align}
where $U_{\gamma}$ satisfies,
\begin{align}
U_{\gamma} \gamma U^{\dagger}_{\gamma} = \left[ \begin{array}{cc} -I_2 & O_2 \\ O_2 & I_2 \end{array} \right].
\end{align}
As shown in Ref.~[\onlinecite{sato_11}], the winding number can be further simplified to:
\begin{align}
\begin{split}
&w(k_y) = -\frac{1}{2} \sum_{C(p_x,k_y)=0} \mathrm{sgn}[\partial_{p_x}C(p_x,k_y)] \mathrm{sgn}[R(p_x,k_y)],\\
&R(p_x,k_y) = \mathrm{Re}[\mathrm{det}\;q(p_x,k_y)] = \xi_+(p_x,k_y) \xi_-(p_x,k_y) + \Delta^2,\\
&C(p_x,k_y) = \mathrm{Im}[\mathrm{det}\;q(p_x,k_y)] = 2 t \Delta \sin \frac{p_x}{2},\\
&\xi_{\pm}(p_x,k_y)=-2t \cos k_y -\mu \pm 2 \sqrt{t^2 \sin^2 \frac{p_x}{2} + t_J^2 \left( \cos \frac{p_x}{2} + \cos k_y \right)^2},
\end{split}
\end{align}
where the summation $\sum_{C(p_x,k_y)=0}$ is taken for $p_x$ satisfying $C(p_x,k_y)=0$ with a fixed $k_y$.
Finally, the winding number is given by:
\begin{align}
\begin{split}
w_+(k_y) = \frac{1}{2}\left\{\mathrm{sgn}[\xi_+(2\pi,k_y) \xi_-(2\pi,k_y) + \Delta^2] - \mathrm{sgn}[\xi_+(0,k_y) \xi_-(0,k_y) + \Delta^2]  \right\},
\end{split}
\end{align}
which is equivalent to Eq.~(14) in the main text.
For $\mu=0$, we observe the particular relations: $\xi_{\pm}(0,\pi - k_y) = -\xi_{\mp}(2\pi,k_y)$ and $\xi_{\pm}(2 \pi, \pi - k_y) = -\xi_{\mp}(0,k_y)$, which lead to $w_+(k_y) = - w_+(\pi - k_y)$.
Consequently, the Atiyah--Singer index, $\mathbb{Z}=\sum^{\prime}_{k_y} w_{+}(k_y)$, becomes identically zero when $\mu=0$, as numerically demonstrated in Fig.~3(c) of the main text.

Next, we discuss the winding number $w_-(k_y)$ associated with class DIII.
According to Ref.~[\onlinecite{ikegaya_18}], the winding number $w_-(k_y)$ must be an odd function of $k_y$, i.e., $w_-(k_y)=-w_-(-k_y)$.
However, the BdG Hamiltonian of the present system satisfies $H(k_x,k_y)=H(k_x,-k_y)$.
Therefore, we can immediately conclude that $w_-(k_y)=0$ identically.

\subsection{Topological phase boundaries}
\begin{figure}[t]
\begin{center}
\includegraphics[width=0.4\textwidth]{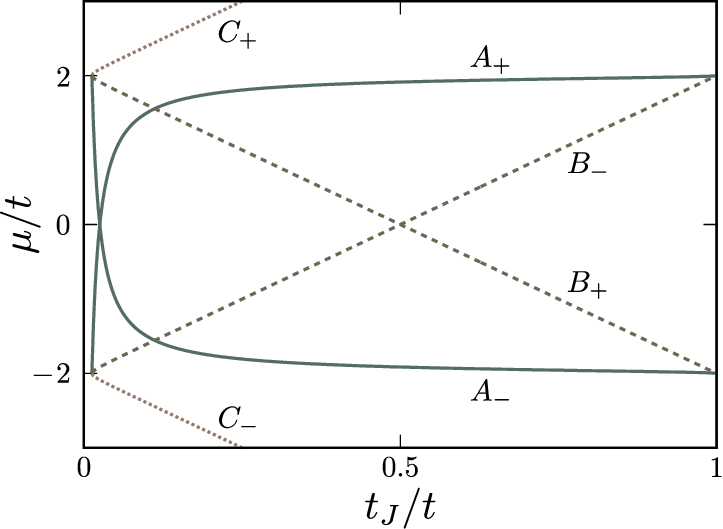}
\caption{Phase diagram for the number of superconducting gap nodes, where we choose $\Delta=0.05t$.
The number of gap nodes in each region is explicitly mentioned in the main text.}
\label{fig:figure_sp1}
\end{center}
\end{figure}

In this section, we present the explicit functions that define the topological phase boundaries shown in Fig.~2 of the main text.
The superconducting gap nodes appear at momenta where $\mathrm{det}[ H(\boldsymbol{k})]=0$, which is equivalent to the condition:
\begin{align}
\mathrm{det}[q(p_x,k_y)] = [\xi_+(p_x,k_y) \xi_-(p_x,k_y) + \Delta^2] + i [2 t \Delta \sin (p_x/2)] = 0.
\end{align}
Thus, we obtain the superconducting gap nodes at $(p_x,k_y)=(0, \pm k_n)$ and $(2 \pi, \pm k_n)$, where $k_n$ satisfies the following condition:
\begin{align}
\xi_+(0,k_n) \xi_-(0,k_n) + \Delta^2 = 0 \quad \mathrm{or} \quad \xi_+(2\pi,k_n) \xi_-(2\pi,k_n) + \Delta^2 = 0.
\label{eq:node_condition}
\end{align}
By performing several calculations, we obtain the phase diagram with respect to the number of $k_n$ satisfying Eq.~(\ref{eq:node_condition}), as shown in Fig.~\ref{fig:figure_sp1}.
The solid lines $A_{\pm}$, which are defined for $|\mu|\leq \sqrt{4t^2-\Delta^2}$, are given by:
\begin{align}
\mu = \pm 2t \mp \sqrt{\frac{t^2-t_J^2}{t_J^2}}\Delta.
\end{align}
The dashed lines $B_{\pm}$, which are also defined for $|\mu|\leq \sqrt{4t^2-\Delta^2}$, are represent by:
\begin{align}
\mu = \pm 2t \mp \sqrt{16t_J^2-\Delta^2}.
\end{align}
The dotted lines $C_{\pm}$, defined for $\mu >\sqrt{4t^2-\Delta^2}$ and $\mu < -\sqrt{4t^2-\Delta^2}$, are given by
\begin{align}
\mu = \pm 2t \pm \sqrt{16t_J^2-\Delta^2}.
\end{align}
The number of gap nodes varies as the parameters intersect these topological phase boundaries, while the number of nodes in each region is specified in the main text.

\subsection{Winding number for accidental band crossings}
In this section, we discuss the additional band crossings that appear in the energy spectra shown in Figs.~3(b) and 3(c) of the main text.
For $k_y = \pm k_F$ with $k_F= \arccos(-\mu/2t)$, the single-particle Hamiltonian accidentally preserves chiral symmetry:
\begin{align}
\gamma_N h(k_x,\pm k_F) \gamma_N^{-1} = - h(k_x,\pm k_F), \quad \gamma_N = \left[ \begin{array}{cc} s_x & O_2 \\ O_2 & -s_x \end{array}\right],
\end{align}
which leads to the additional chiral symmetry in the BdG Hamiltonian at $k_y = \pm k_F$,
\begin{align}
\Gamma_N  H(k_x,\pm k_F) \Gamma_N^{-1} = -H(k_x,\pm k_F), \quad \Gamma_N =\left[ \begin{array}{cc} \gamma_N & O_4 \\ O_4 & \gamma_N \end{array} \right].
\end{align}
Using this chiral symmetry, we can define the winding number applicable only at $k_y = \pm k_F$:
\begin{align}
\begin{split}
w_N (\pm k_F) &= \frac{i}{4 \pi} \int^{\pi}_0 dk_x \mathrm{Tr}[\Gamma_N H^{-1}(k_x,\pm k_F) \partial_{k_x} H(k_x,\pm k_F)]\\
&=\frac{i}{4 \pi} \int^{\pi}_0 dk_x \mathrm{Tr}[A_1\Gamma_N A_1^{\dagger} A_1 H^{-1}(k_x,\pm k_F) A_1^{\dagger} \partial_{k_x} A_1 H(k_x,\pm k_F) A_1^{\dagger}]\\
&=\frac{i}{4 \pi}\sum_{s=\pm} \int^{\pi}_0 dk_x \mathrm{Tr}[\gamma^{\prime}_N \{H^{\prime}_s\}^{-1}(k_x,\pm k_F) \partial_{k_x} H^{\prime}_s (k_x,\pm k_F) ],
\end{split}
\end{align}
where
\begin{align}
\gamma^{\prime}_N = \left[ \begin{array}{cc} -\sigma_x & O_2 \\ O_2 & -\sigma_x \end{array} \right].
\end{align}
The chiral symmetry operator $\gamma^{\prime}_N$ is diagonalized as:
\begin{align}
U_N \gamma^{\prime}_N U_N^{-1} =  \left[ \begin{array}{cc} -I_2 & O_2 \\ O_2 & I_2 \end{array} \right], \quad
U_N = \frac{1}{2}\left[ \begin{array}{cccc} 1& 1& 0& 0 \\ 0& 0& 1& 1 \\ 1& -1& 0& 0 \\ 0& 0& 1& -1 \end{array} \right].
\end{align}
Using the unitary operator $U_N$, we obtain the chiral basis Hamiltonian with respect to $\gamma^{\prime}_N$,
\begin{align}
\begin{split}
&U_{N} H^{\prime}_{\pm}(k_x,\pm k_F) U^{\dagger}_{N} = \left[ \begin{array}{cc} O_2 & q_{N,\pm} (k_x) \\ q_{N,\pm}^{\dagger}(k_x) & O_2 \end{array} \right],\\
&q_{N,\pm} (k_x) = \pm \left[-2t \cos \frac{k_x}{2} + 2it_J \left\{ \sin \frac{k_x}{2} \pm \frac{\mu}{2t} \right\}\right] I_2 - i \Delta \sigma_y.
\end{split}
\end{align}
According to Ref.~[\onlinecite{sato_11}], we can calculate the winding number simply by
\begin{align}
\begin{split}
&w_N(\pm k_F) = -\frac{1}{2} \sum_{s=\pm} \sum_{C_{N,s}(k_x)=0} \mathrm{sgn}[\partial_{k_x}C_{N,s}(k_x)] \mathrm{sgn}[R_{N,s}(k_x)],\\
&R_{N,s}(k_x) = \mathrm{Re}[\mathrm{det}\;q_{N,s}(k_x)],
C_{N,s}(k_x) = \mathrm{Im}[\mathrm{det}\;q_{N,s}(k_x)],
\end{split}
\end{align}
where the summation $\sum_{C_{N,s}(k_x)=0}$ is taken for $k_x$ satisfying $C_{N,s}(k_x)=0$.
For simplicity, we now consider the limit as $\Delta \rightarrow 0$.
In this limit, the winding number is further simplified to:
\begin{align}
w_N(\pm k_F) = \frac{1}{2}\sum_{s=\pm} \left\{ \mathrm{sgn}\left[1 + s \frac{\mu}{2t} \right] + \mathrm{sgn}\left[1 - \left(\frac{\mu}{2t}\right)^2 \right] \right\}.
\end{align}
Since $|\mu/2t| <1$, which ensures $k_F= \arccos(-\mu/2t) \in \mathbb{R}$, we find that $w_N(\pm k_F)=+2$, indicating the presence of the two zero-energy edge states at $k_y = \pm k_F$.
Consequently, we conclude that the band crossings observed in Figs.~3(b) and 3(c) of the main text are due to the appearance of additional edge states,
which are protected by the accidental normal-state chiral symmetry $\Gamma_N$.

\section{Majorana Flat Bands in Various Topological Phases}
\begin{figure*}[t]
\begin{center}
\includegraphics[width=0.8\textwidth]{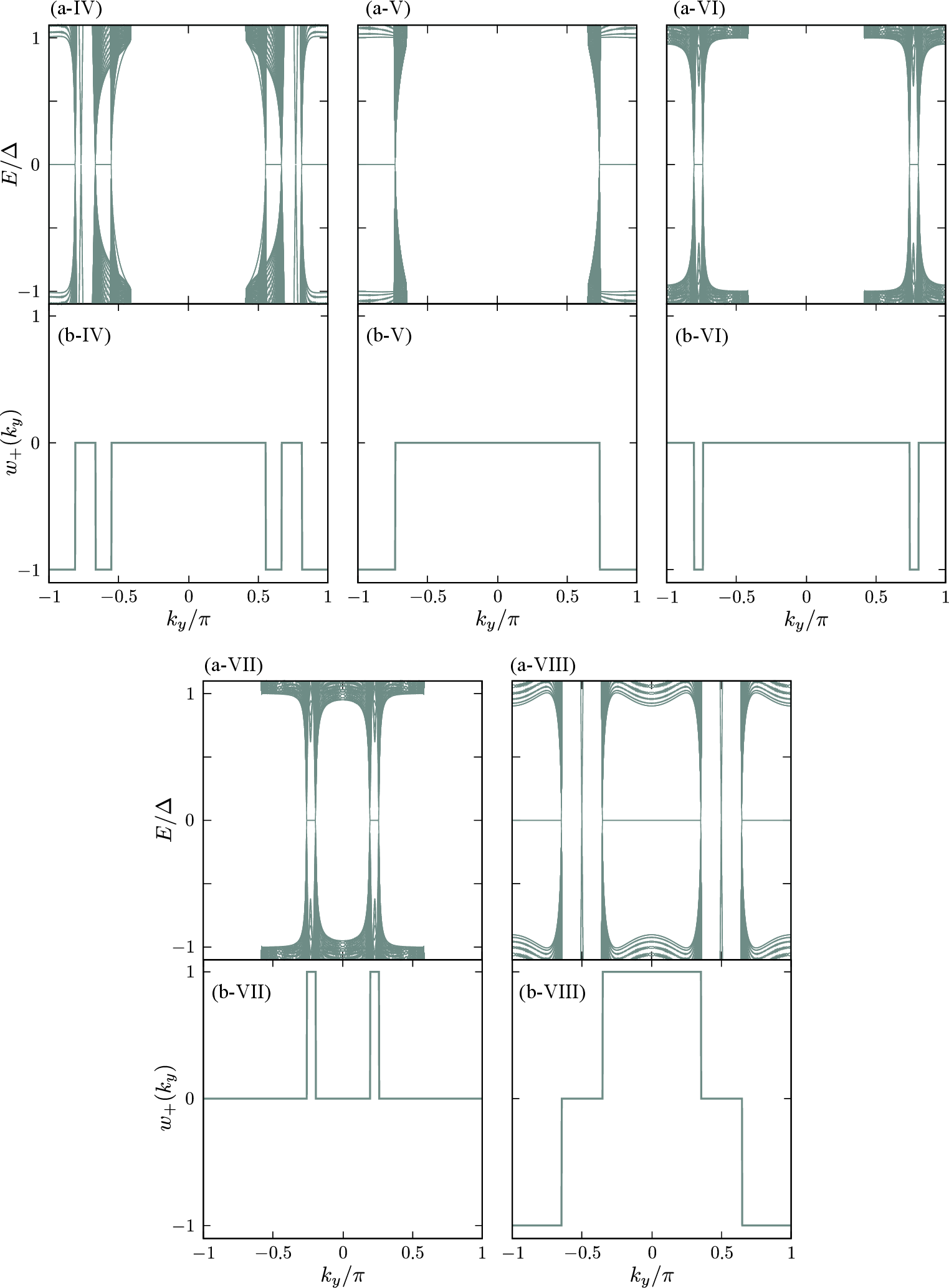}
\caption{Energy spectra and winding numbers of the nodal topological phases IV--VIII.
The upper panels display the energy spectra for these phases, while the corresponding winding numbers $w_+(k_y)$ are shown in the lower panels.}
\label{fig:figure_sp2}
\end{center}
\end{figure*}
In this section, we discuss the energy spectra for the nodal topological phases IV--VIII, as shown in the upper panels (a-IV)--(a-VIII) in Fig.~\ref{fig:figure_sp2}.
The lower panels (b-IV)--(b-VIII) display the corresponding winding number $w_+(k_y)$.
To describe phases, IV--VIII, we choose the following parameters: $(\mu/t,t_J/t)=(1.5,0.5)$, $(3,0.5)$, $(1.5,0.04)$, $(-1.5,0.04)$, and $(0, 0.8)$, respectively.
The pair potential is fixed at $\Delta=0.05t$.
We confirm the appearance of the Majorana flat-band states (MFBSs) in the momentum ranges where $w_+(k_y) \neq 0$.
For phases IV and V, MFBSs are found around $k_y = \pm \pi$, where the corresponding winding numbers are negative.
In phases VI and VII, the momentum ranges hosting MFBSs are very narrow, while these phases vanish in the limit of $\Delta/t \rightarrow 0$.
For phase VIII, we find $\mathbb{Z}=\sum^{\prime}_{k_y} w_{+}(k_y) = 0$, since we choose $\mu=0$ to describe this phase.
\end{document}